\documentclass{article}
\pdfoutput=1
\usepackage[T1]{fontenc}
\usepackage[latin9]{inputenc}
\usepackage{amsmath}
\usepackage{amssymb}
\usepackage{graphicx}
\usepackage{CJK}
\usepackage{amsmath}
\usepackage{cases}
\usepackage{indentfirst}
\usepackage{cite}

\begin{document}
\title{Two-distance minimal dominating set problem studied by statistical mechanics and simulated annealing}
\author{\emph{Yusupjan Habibulla} \\School of Physics and Technology, Xinjiang University, Sheng-Li Road 666,\\  Urumqi 830046, China\\ \emph{Shao-Meng Qin}\\College of Science, Civil Aviation University of China\\Tianjin 300300, China}

 \maketitle
\begin{abstract}
The L-distance (especially the 2-distance) minimal dominating set (MDS) problem is widely considered in various dominating set problems. Recently, we studied the regular dominating set problem using the cavity method and developed two algorithms (belief propagation decimation and survey propagation decimation) to estimate the solution of a given graph, resulting in very good estimations of the minimal dominating size. This paper describes the development of spin glass theory to study the 2-distance MDS problem. First, we show that the belief propagation equation does not converge when the inverse temperature $\beta$ is greater than a certain threshold value on both regular and Erdos--Renyi random networks. Second, we find that the entropy density is equal to zero at finite inverse temperature on regular random graphs when the node degree is from 3--9, and on Erdos--Renyi random networks when the node degree is from 4.2--10.4; the entropy density is positive in all other cases. This result is proved using a dynamical simulated annealing process. Third, the results of replica symmetry theory are shown to be in agreement with those of the belief propagation algorithm, and the results of the belief propagation decimation algorithm are found to be better than those of the greedy heuristic algorithm. 
\end{abstract}
\section{Introduction}
There is a close relationship between the minimal dominating set (MDS) and the L-distance minimal dominating set (L-MDS). The idea behind the MDS is to construct the node set of smallest size such that any node of the network is either in this set or is adjacent to at least one node of this set. The allocation of network resources that satisfies a given service with the least use of resources is a frequent problem in the field of communication networks. We are interested in the problem of computing the minimum set of covered nodes (i.e., servers) such that every node is covered or has at least one covered neighbor node at a distance of at most $L$ (i.e., the L-MDS problem), where the distance between two nodes in the graph is the minimum number of hops necessary to move from one to the other. Each server then provides a service to or monitors those nodes within a distance of $L$. Consider a simple network $W$ formed by $N$ nodes and $M$ undirected links, where each link connects two different nodes. The edge density $\alpha$ is simply defined as $\alpha \equiv M/N$, and the node degree $C$ is defined as $C\equiv 2\times\alpha$. The 2-distance MDS $\gamma $ of the given network $W$ is the smallest set of nodes such that all nodes of the network either belong to this set or have at least one neighbor node within 2-distance that belongs to this set. If node $i$ belongs to the 2-distance MDS, we say that $i$ is covered. In Fig. 1, the green nodes (i.e., covered nodes) construct a 2-distance MDS for the given small graph. If node $i$ belongs to the 2-distance MDS or at least one neighbor node within 2-distance is covered, we say that $i$ is observed. In the figure, each blue node has at least one covered neighbor node within 2-distance, so all of the blue nodes are observed.\\
Many new types of dominating set problem have been studied by mathematicians and computer scientists in recent years. For example, Sehrawat et al. studied double dominating sets \cite{1,5}, and Jena et al. 
\begin{figure}[!htbp]
\centering
\includegraphics[width=10cm,height=10cm]{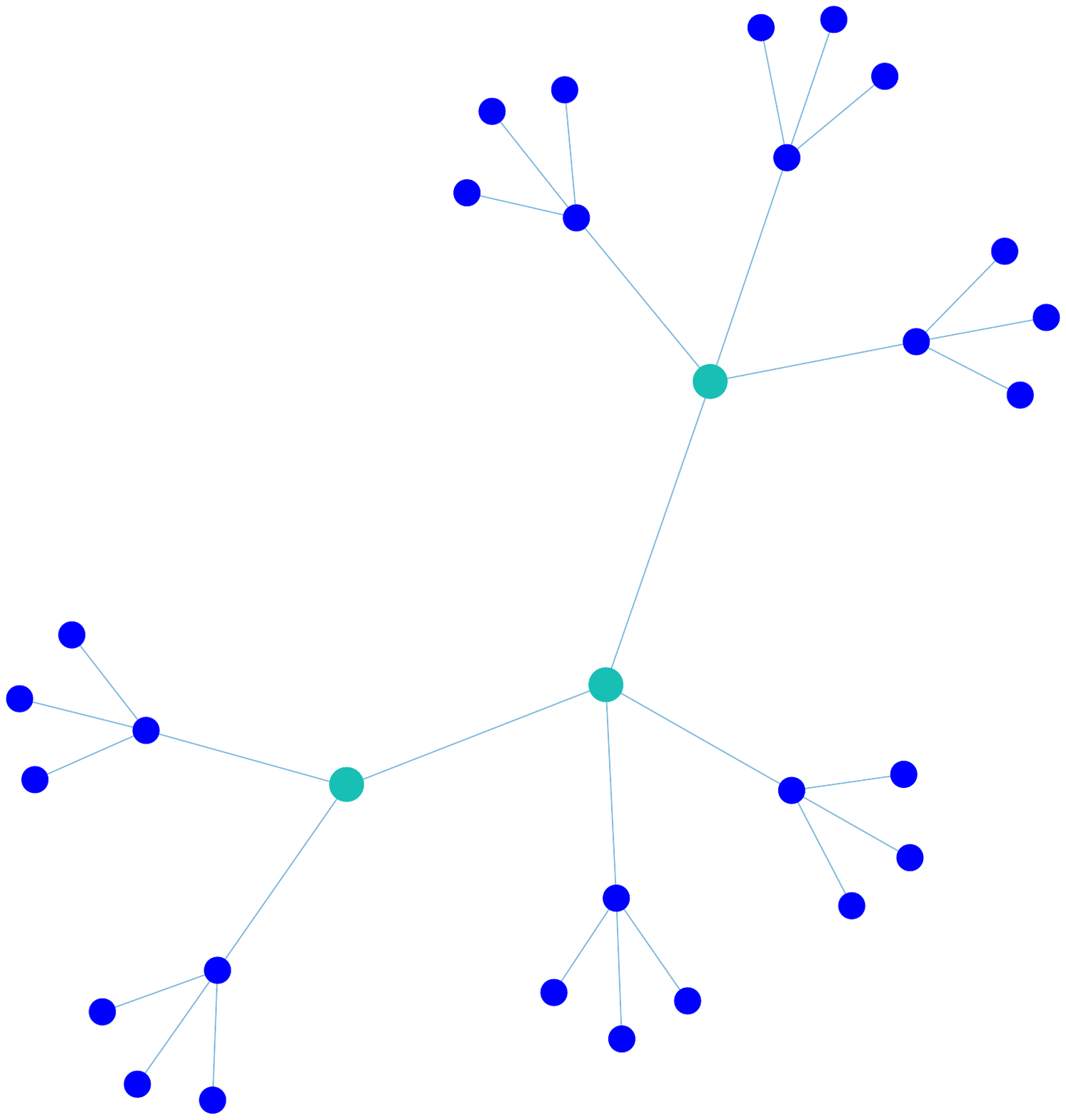}
\caption{Example of a 2-distance MDS in a small network with N = 31 nodes and M = 30 links. Green nodes are covered, while blue nodes are empty but observed. The three covered nodes form a 2-distance MDS for this network.}
\end{figure}
studied a generalized liar's dominating set \cite{2} called the distance-d $(m,l)$-liar's dominating set. This is a subset $D\subseteq V$ such that $(i)$ each vertex in V is distance-d dominated by at least $m$ vertices in D, and $(ii)$ each pair of distinct vertices in V is distance-d dominated by at least $l$ vertices in D, where $m<l$. They proved that this problem is $NP$-complete. Alzoubi et al. studied weakly connected dominating sets~\cite{3}, in which if set $D$ is a weakly connected dominating set, then $(i)$ D is a minimal dominating set, and $(ii)$ the nodes in $D$ are connected at least in 2-distance. The authors provided two algorithms to solve this problem. Wu et al. studied extended dominating sets~\cite{4} using cooperative communication, in which each node is covered by either one or several 2-distance neighbors.  They proposed several heuristic algorithms to construct a minimal extended dominating set. Lamprou et al. proposed budgeted connected domination and budgeted edge-vertex domination~\cite{6}. The budgeted connected dominating set has applications in ad hoc wireless networks, while the budgeted edge-vertex dominating set considers scenarios in which resources must be positioned on the links of a network to dominate network nodes. Iqbal et al. studied the 2-distance paired domination of the flower graph~\cite{7}, which is a 2-distance dominating set,  and showed that its induced subgraph has a perfect matching. Beggas et al. studied the [1,2]-dominating set~\cite{8}, in which every vertex not in the [1,2]-dominating set  has at least one and at most two neighbors in this set. Other studies have considered ($\sigma , \rho $) dominating sets~\cite{9}, convex dominating sets~\cite{10}, $k$-tuple dominating sets~\cite{11}, directed edge dominating sets~\cite{12}, and $k$-connected $m$-dominating sets~\cite{13}. The dominating set problem is increasingly applicable to new scientific and technical fields~\cite{14,15,16,17,18,19,20,21,22,23}. \\
Although there are many types of dominating set, most of them are of three types: regular dominating set, connected dominating set, or L-distance dominating set. Some work has focused purely on the L-distance dominating set problem~\cite{24,25} and the connected dominating set problem~\cite{26,27}. The statistical physics of spin glass systems have been widely applied to optimization problems, such as in the minimal vertex cover problem~\cite{28,29}, the minimum feedback vertex set problem~\cite{30,31}, and to satisfiability problems, such as in K-SAT~\cite{32,33,34}, XOR-SAT~\cite{35,36}, and the MDS problem~\cite{37,38,39,40}. Recently, we used statistical physics to study the regular MDS problem.  We introduced belief propagation decimation (BPD), warning propagation, and survey propagation decimation algorithms to obtain the MDS, and showed that our algorithms produce solutions that are very close to optimal in a very short computation time. The solution space has a condensation transition and a cluster transition on undirected regular random (RR) graphs, but has only one transition on undirected Erdos--Renyi (ER) random graphs and on both directed ER random and RR graphs. In this paper, we describe the further use of statistical physics to study the 2-distance MDS. The ground-state energy appears when the entropy density is equal to zero on the undirected ER random graph, and the entropy is always positive when the mean degree is outside the range 4.2--10.4. The ground-state energy still appears when the entropy is equal to zero on the undirected RR random graph, which occurs when the node degree is in the range 3--9. We use population dynamics, BPD, and greedy heuristic algorithms to estimate the 2-distance MDS. The results demonstrate that the population dynamics and BPD results are always better than those of the greedy heuristic algorithm on single ER and RR random graphs.\\
The remainder of this paper is organized as follows. In Section 2, we discuss the concept of replica symmetry (RS) theory for the 2-distance MDS problem and present the belief propagation (BP) equation and corresponding thermodynamic quantities. In Section 3, we introduce the BPD algorithm and greedy algorithm for the 2-distance MDS problem, and derive the BP equation and marginal probability equation for the different node state conditions. We also construct the proper BPD process for estimating the 2-distance MDS. In Section 4, we introduce the rejection-free simulated annealing process for the 2-distance MDS problem. Finally, Section 5 presents the conclusions from this study and summarizes our results. 
\section{Replica symmetry}
In this section, we introduce the mean field theory for the 2-distance MDS problem. The energy function of the 2-distance MDS problem cannot be written in the standard manner, but we can write the partition function in the standard form. Depending on the RS mean field theory of statistical physics, we can write the partition function $Z$ as
\begin{equation}
Z=\sum_{\underline{c}}\prod_{i\in W}{e^{-\beta \delta_{ c_{i}}^{0}}}[1-(1-\delta_{ c_{i}}^{0})\prod_{j\in\partial i}(1- \delta_{ c_{j}}^{c_{i}-1})-\Theta(\sum\limits_{j\in\partial i}(\delta_{ c_{j}}^{c_{i}+2}+\delta_{ c_{j}}^{c_{i}-2})-1)],
\end{equation}
where $\underline{ c}\equiv( c_{1}, c_{2}, \ldots, c_{n})$ denotes one of the possible configurations, $\delta_{m}^{n}=1$ if $m = n$ and $\delta_{m}^{n}=0$ otherwise, and $\Theta(x)=1$ if $x\geq 0$ and $\Theta(x)=0$ otherwise. $ c_{i}=0$ if node $i$ is covered, $ c_{i}=1,2$ if node $i$ is not covered. If node $i$ is in state $c_{i}=0$, then the neighbor nodes can only take state $c_{k}=0$ or $c_{k}=1$, with state $c_{k}=2$ forbidden.  If node $i$ is in state $c_{i}=1$, then the neighbor nodes may take any state $c_{k}=0$, $c_{k}=1$, or $c_{k}=2$, but at least one neighbor must be covered. If node $i$ is in state $c_{i}=2$, then the neighbor nodes may only take state $c_{k}=1$ or $c_{k}=2$, but at least one 2-distance neighbor must be covered, and state $c_{k}=0$ is forbidden. Therefore, the partition function only takes into account the 2-distance dominating set (DS), and as $\beta\rightarrow\infty$, the contribution is exclusively from the 2-distance MDS configurations.\\

RS mean field theories, such as the Bethe--Peierls approximation~\cite{41} and partition function expansion~\cite{42,43}, can solve the above spin glass model. The results of the Bethe--Peierls approximation are in perfect agreement with those of the partition function expansion algorithm, but the Bethe--Peierls approximation theory equation has a simpler form. Thus, we introduce the Bethe--Peierls approximation equation. We set a cavity message $p_{i\rightarrow j}^{(c_{i},c_{j})}$ on each edge, which must satisfy the BP equation 

\begin{equation}
p_{i\rightarrow j}^{(c_{i},c_{j})}=\frac{e^{-\beta \delta_{ c_{i}}^{0}}\prod\limits_{k\in\partial i\backslash j}\sum\limits_{c_{k}\in A}p_{k\rightarrow i}^{( c_{k}, c_{i})}-(1-\delta_{ c_{i}}^{0})(\delta_{ c_{j}}^{c_{i}}+\delta_{ c_{j}}^{c_{i}+1})\prod\limits_{k\in\partial i\backslash j}\sum\limits_{c_{k}\geq c_{i}}p_{k\rightarrow i}^{(c_{k},c_{i})}}{\sum\limits_{\acute{ c}_{i},\acute{ c}_{j}}e^{-\beta \delta_{\acute c_{i}}^{0}}\prod\limits_{k\in\partial i\backslash j}\sum\limits_{\acute c_{k}\in A}p_{k\rightarrow i}^{( \acute c_{k}, \acute c_{i})}-(1-\delta_{ \acute c_{i}}^{0})(\delta_{ \acute c_{j}}^{\acute c_{i}}+\delta_{ \acute c_{j}}^{\acute c_{i}+1})\prod\limits_{k\in\partial i\backslash j}\sum\limits_{\acute c_{k}\geq \acute c_{i}}p_{k\rightarrow i}^{(\acute c_{k},\acute c_{i})}},
\end{equation}

where $\partial i\backslash j$ denotes the subset obtained by deleting node $j$ from set $\partial i$. The cavity message $p_{i\rightarrow j}^{(c_{i},c_{j})}$ represents the joint probability that node $i$ is in occupation state $c_{i}$ and its adjacent node $j$ is in occupation state $c_{j}$ when the constraint of node $j$ is not considered.  Set $A$ represents the possible states of $c_{k}$. The marginal probability $p_{i}^{c}$ of node $i$ is expressed as
\begin{equation}
p_{i}^{c}=\frac{e^{-\beta \delta_{c}^{0}}\prod\limits_{j\in\partial i}\sum\limits_{c_{j}\in A}p_{j\rightarrow i}^{( c_{j}, c)}-(1-\delta_{ c}^{0})\prod\limits_{j\in\partial i}\sum\limits_{c_{j}\geq c}p_{j\rightarrow i}^{(c_{j},c)}}{\sum\limits_{c_{i}}e^{-\beta \delta_{c_{i}}^{0}}\prod\limits_{j\in\partial i}\sum\limits_{c_{j}\in A}p_{j\rightarrow i}^{(c_{j}, c_{i})}-(1-\delta_{ c_{i}}^{0})\prod\limits_{j\in\partial i}\sum\limits_{c_{j}\geq c_{i}}p_{j\rightarrow i}^{(c_{j},c_{i})}}.
\end{equation}

The messages $p_{j\rightarrow i}^{( c_{j}, c)}$ are converged messages, that is, the marginal probability is calculated after the BP equation converges. $p_{i}^{0}$ denotes the probability that node $i$ is covered, $p_{i}^{1}$ denotes the probability that node $i$ has at least one covered neighbor, and $p_{i}^{2}$ denotes the probability that node $i$ has at least one covered 2-distance neighbor.

The total free energy $F_{0}=\frac{1}{\beta}\ln Z$ is determined by the total partition function. We use the Bethe-Peierls approximation to express the total free energy as  
\begin{equation}
F_{0}=\sum_{i=1}^{N}F_{i}-\sum_{(i,j)\in G}F_{(i,j)},
\end{equation}

where $F_{i}$ denotes the free energy of node $i$ and $F_{(i,j)}$ denotes the free energy of edge $(i,j)$. Thus, the total free energy can be expressed as the sum of the free energy of all nodes and edges. $F_{i}$ and $F_{(i,j)}$ can be expressed as

\begin{equation}
F_{i}=-\frac{1}{\beta}\ln[\sum\limits_{c_{i}}e^{-\beta \delta_{c_{i}}^{0}}\prod\limits_{j\in\partial i}\sum\limits_{c_{j}\in A}p_{j\rightarrow i}^{(c_{j}, c_{i})}-(1-\delta_{ c_{i}}^{0})\prod\limits_{j\in\partial i}\sum\limits_{c_{j}\geq c_{i}}p_{j\rightarrow i}^{(c_{j},c_{i})}],
\end{equation}

\begin{equation}
F_{(i,j)}=-\frac{1}{\beta}\ln[\sum_{ c_{i}, c_{j}\in A}p_{i\rightarrow j}^{( c_{i}, c_{j})}p_{j\rightarrow i}^{( c_{j}, c_{i})}].
\end{equation}

The total free energy is equal to the sum of the free energy contributions from each node. From the expression of $F_{i}$, we can see that the free energy of each node includes the contribution of the node and of the edges between the node and all neighboring nodes, so each edge is considered twice. Thus, the free energy contribution of all edges must be subtracted, which is why we have the expression given in Eq. (4) for the total free energy. The BP equation is iterated until it converges to one stable point, and then the mean free energy $f\equiv F/N$ and energy density $E=1/N\sum_{i}p_{i}^{0}$ are calculated using Eqs.~(3) and (4). The entropy density is calculated as $s=\beta(E-f)$.

\begin{figure}[!htb]
  \centering
  \includegraphics[width=12cm,height=7cm]{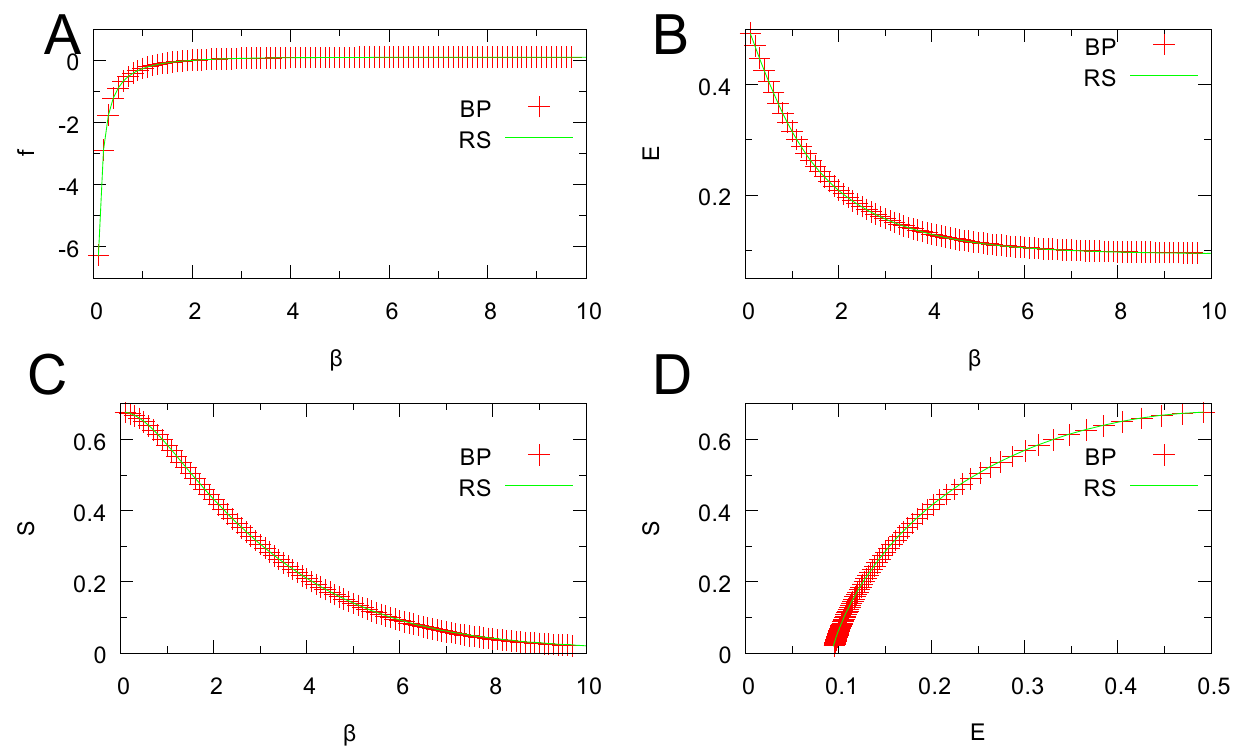}
  \caption{RS and BP results for the 2-distance MDS problem on an ER random graph with node degree $c=4$ and $N=10000$ using BP and population dynamics. In subgraphs A, B, C, the $x$-axis denotes the inverse temperature $\beta $ and the $y$-axis denotes the thermodynamic quantities. In subgraph D, the $x$-axis denotes the energy density and the $y$-axis denotes the entropy density.}
\end{figure}

From Eq. (2) we can see that each node has seven probability messages (namely $p_{i\rightarrow j}^{(0,0)}$, $p_{i\rightarrow j}^{(0,1)}$, $p_{i\rightarrow j}^{(1,0)}$, $p_{i\rightarrow j}^{(1,1)}$, $p_{i\rightarrow j}^{(1,2)}$, $p_{i\rightarrow j}^{(2,1)}$, and $p_{i\rightarrow j}^{(2,2)}$), where $p_{i\rightarrow j}^{(0,0)}=p_{i\rightarrow j}^{(0,1)}$ and $p_{i\rightarrow j}^{(1,1)}=p_{i\rightarrow j}^{(1,2)}$. Therefore, we use five messages (namely $p_{i\rightarrow j}^{(0,0)}$, $p_{i\rightarrow j}^{(1,0)}$, $p_{i\rightarrow j}^{(1,1)}$, $p_{i\rightarrow j}^{(2,1)}$, and $p_{i\rightarrow j}^{(2,2)}$) to describe the probability distribution $p_{i\rightarrow j}^{(c_{i},c_{j})}$, so the normalization condition can be written as
\begin{equation}
2p_{i\rightarrow j}^{(0,0)}+p_{i\rightarrow j}^{(1,0)}+2p_{i\rightarrow j}^{(1,1)}+p_{i\rightarrow j}^{(2,1)}+p_{i\rightarrow j}^{(2,2)}=1.
\end{equation}
The message pair $p_{i\rightarrow j}^{(c_{i},c_{j})}$ and $p_{j\rightarrow i}^{(c_{j},c_{i})}$ is randomly initialized on each edge $(i,j)$ between nodes $i$ and $j$, and all message pairs are identical to each other. The BP equation updates the message on each edge of the given network until it converges to a stable point. We calculate the difference of every message on each edge by $\Delta_{i\rightarrow j}^{(c_{i},c_{j})}=|p_{i\rightarrow j}^{(c_{i},c_{j})}(t)-p_{i\rightarrow j}^{(c_{i},c_{j})}(t-1)|$, where $p_{i\rightarrow j}^{(c_{i},c_{j})}(t)$ is the updated message at the $t$-th BP step and $p_{i\rightarrow j}^{(c_{i},c_{j})}(t-1)$ is the old message at step $(t-1)$, and then return the biggest difference of messages. Finally, we select the biggest difference among all edges and compare this with a pre-specified threshold value (e.g., $10^{-10}$ or even smaller). The BP iterations stop if the biggest difference is less than the pre-specified threshold value; otherwise, they continue. We perform the BP iterations at a given value of the inverse temperature $\beta$ until a fixed point is found, and then calculate thermodynamic quantities such as the free energy $f$, mean energy $\omega$, and entropy $s$. The BP iteration process produces different results on ER random networks and RR networks. Figure 2 shows the results for an ER random network with mean node degree 4 and $10^{4}$ nodes, while Fig. 3 shows the results for an RR network with mean degree 3 and $10^{4}$ nodes.\\
For RR networks with a node degree outside the range 3--9 and ER random networks with a mean node degree outside the range 4.2--10.4, we find that the BP equation does not converge to a stable point when the inverse temperature is above a certain threshold value. For example, the BP equation does not converge when the inverse temperature $\beta > 9.7$ on ER random networks with mean node degree 4. This phenomenon indicates that the system will be in the spin glass phase when the inverse temperature is sufficiently large at some particular node degree. In Tables 1 and 2, we present the ground state energies and corresponding inverse temperatures of RR networks and ER random networks. We can see that the inverse temperature $\beta_{d}$ first decreases and then increases as the node degree rises for ER random networks, while it increases monotonically with the node degree in RR networks. The energy density is always a monotonically decreasing function of the node degree. \\
We compute the population dynamics using Eqs. (2)--(4) to obtain the ensemble-averaged results in the same way as the RS population dynamics for the MDS problem \cite{37}; thus, the details are omitted here. Roughly speaking, we first create a long array $A$ of $N$ elements, each element representing a probability distribution $p_{i\rightarrow j}^{(c_{i},c_{j})}$ in the form of a five-dimensional vector. Then, we update the array sufficiently many times (e.g., each element of this array is updated 200 times on average) using Eq. (2), and then compute the thermodynamic quantities $f,\omega$, and $s$ using Eqs. (3) and (4). \\
In Fig. 2, we compare the results of the RS population dynamics with the BP algorithm on a single ER random graph with $C=4$. The ensemble-averaged results are in agreement with the results of the BP algorithm. In some random network systems, the entropy density becomes negative when the energy density $\omega$ is less than a certain threshold value, $\omega_{0}$ (see Fig. 3(D)), indicating that there is no 2-distance dominating set with relative size below $\omega_{0}$. We take $\omega_{0}$ as the ensemble-averaged 2-distance MDS size (see Table 1 for ER random networks and Table 2 for RR networks). In some other random network systems (Fig. 2), the entropy density approaches a positive limiting value as the energy density reaches a threshold $\omega_{0}$ (see Fig. 2(D)). We take $\omega_{0}$ as the ensemble-averaged 2-distance MDS size of this random network. 
\begin{table}[!hbp]
\caption{Inverse temperature $\beta_{d}$ and corresponding energy density for ER random graphs when the entropy density is equal to zero.}
\begin{tabular}{p{0.9cm}p{0.9cm}p{0.9cm}p{0.9cm}p{0.9cm}p{0.9cm}p{0.9cm}p{0.9cm}p{0.9cm}}
\hline
C & 4.2 & 5 & 6 & 7 & 8 & 9 & 10 &10.4\\
\hline
$\beta_{d}\approx$ & 15.25 & 11.05 & 10.45 & 10.75&11.25&12.05&13.55&15.85\\
\hline
$E_{min}\approx$ & 0.0846 & 0.0612 & 0.0440 & 0.0337&0.0270&0.0223&0.0189&0.0176\\
\hline
\end{tabular}
\end{table}

\begin{figure}[!htb]
  \centering
  \includegraphics[width=12cm,height=7cm]{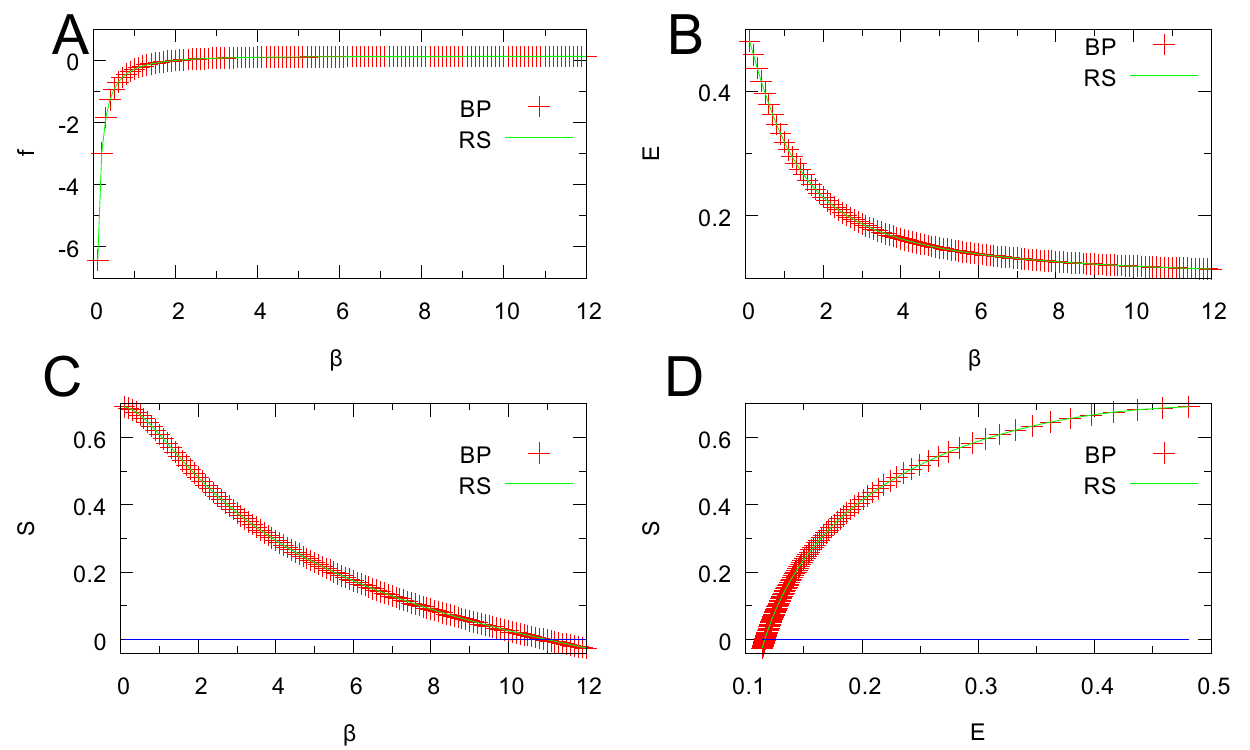}
  \caption{RS and BP results for the 2-distance MDS problem on an RR random graph with node degree $c=3$ and $N=10^{4}$ using BP and population dynamics. In subgraphs A, B, C, the $x$-axis denotes the inverse temperature $\beta $ and the $y$-axis denotes the thermodynamic quantities. In subgraph D, the $x$-axis denotes the energy density and the $y$-axis denotes the entropy density.}
\end{figure}

\begin{table}[!htb]
\caption{Entropy transition point inverse temperature $\beta_{d}$ and the corresponding energy density for RR graphs.}
\begin{tabular}{p{0.9cm}p{0.9cm}p{0.9cm}p{0.9cm}p{0.9cm}p{0.9cm}p{0.9cm}p{0.9cm}}
\hline
C & 3 & 4 & 5 & 6&7&8&9\\
\hline
$\beta_{d}\approx$ & 10.95 & 11.25 & 12.15 & 13.11&14.05&15.05&15.95\\
\hline
$E_{min}\approx$ & 0.1165 & 0.0795 & 0.0592 & 0.0460&0.0371&0.0305&0.0258 \\
\hline
\end{tabular}
\end{table}

\section{Belief propagation decimation algorithm and greedy algorithm}
In this paper, we examine the use of two algorithms to determine the solution of the given graph: the greedy algorithm and the BPD algorithm. The greedy algorithm is very fast, but it does not guarantee good results. The BPD algorithm is not as fast as the greedy algorithm, but it always provides a good estimation for the 2-distance MDS problem.  
\subsection{Belief propagation decimation}

If node $i$ is unobserved (i.e., it is empty and all neighbor and 2-distance neighbor nodes are not covered), then the output message $p_{i\rightarrow j}$ on link $(i,j)$ between nodes $j$ and $i$ is updated according to Eq. (2). In contrast, if node $i$ is empty but observed and has at least one covered neighbor node, that is, $c_{i}=1$, then this node presents no restriction to the states of all its uncovered neighbors. For such a node $i$, there is no possibility that $c_{i}=2$, and so the cavity message $p_{i\rightarrow j}$ on link $(i,j)$ is then updated according to 

\begin{equation}
p_{i\rightarrow j}^{(c_{i},c_{j})}=\frac{e^{-\beta \delta_{ c_{i}}^{0}}(1-\delta_{c_{i}}^{2})\prod\limits_{k\in\partial i\backslash j}\sum\limits_{c_{k}\in A}p_{k\rightarrow i}^{( c_{k}, c_{i})}}{\sum\limits_{\acute{ c}_{i},\acute{ c}_{j}}e^{-\beta \delta_{\acute c_{i}}^{0}}(1-\delta_{c_{i}}^{2})\prod\limits_{k\in\partial i\backslash j}\sum\limits_{\acute c_{k}\in A}p_{k\rightarrow i}^{( \acute c_{k}, \acute c_{i})}}.
\end{equation}

For node $i (c_{i}=1)$, if at least one neighbor node $j$ is covered, then node $j$ sends the message $p_{j\rightarrow i}^{(0, 0)}=p_{j\rightarrow i}^{(0, 1)}=0.5$ to node $i$.  Thus, $p_{j\rightarrow i}^{(0, 1)}+p_{j\rightarrow i}^{(1, 1)}+p_{j\rightarrow i}^{(2, 1)}=p_{j\rightarrow i}^{(0, 1)}$, so the constraints of node $i$ with respect to all other neighbor nodes are automatically removed.  The marginal probability is calculated by
\begin{equation}
p_{i}^{c}=\frac{e^{-\beta \delta_{c}^{0}}(1-\delta_{c_{i}}^{2})\prod\limits_{j\in\partial i}\sum\limits_{c_{j}\in A}p_{j\rightarrow i}^{( c_{j}, c)}}{\sum\limits_{c_{i}}e^{-\beta \delta_{c_{i}}^{0}}(1-\delta_{c_{i}}^{2})\prod\limits_{j\in\partial i}\sum\limits_{c_{j}\in A}p_{j\rightarrow i}^{(c_{j}, c_{i})}}.
\end{equation}

If node $i$ is empty but observed (i.e., it has no adjacent covered node, but has at least one covered 2-distance neighbor node), this node then presents no restriction to the states of its uncovered 2-distance neighbors. For such a node $i$, the output message $p_{i\rightarrow j}$ on link $(i,j)$ is then updated according to 
\begin{equation}
p_{i\rightarrow j}^{(c_{i},c_{j})}=\frac{e^{-\beta \delta_{ c_{i}}^{0}}\prod\limits_{k\in\partial i\backslash j}\sum\limits_{c_{k}\in A}p_{k\rightarrow i}^{( c_{k}, c_{i})}-(1-\delta_{ c_{i}}^{0}-\delta_{ c_{i}}^{2})(\delta_{ c_{j}}^{c_{i}}+\delta_{ c_{j}}^{c_{i}+1})\prod\limits_{k\in\partial i\backslash j}\sum\limits_{c_{k}\geq c_{i}}p_{k\rightarrow i}^{(c_{k},c_{i})}}{\sum\limits_{\acute{ c}_{i},\acute{ c}_{j}}e^{-\beta \delta_{\acute c_{i}}^{0}}\prod\limits_{k\in\partial i\backslash j}\sum\limits_{\acute c_{k}\in A}p_{k\rightarrow i}^{( \acute c_{k}, \acute c_{i})}-(1-\delta_{ \acute c_{i}}^{0}-\delta_{ \acute c_{i}}^{2})(\delta_{ \acute c_{j}}^{\acute c_{i}}+\delta_{ \acute c_{j}}^{\acute c_{i}+1})\prod\limits_{k\in\partial i\backslash j}\sum\limits_{\acute c_{k}\geq \acute c_{i}}p_{k\rightarrow i}^{(\acute c_{k},\acute c_{i})}}.
\end{equation}

For node $i (c_{i}=2)$, if at least one neighbor node $j$ takes state $c_{j}=1$, then node $j$ sends the message $p_{j\rightarrow i}^{(2, 1)}=p_{j\rightarrow i}^{(2, 2)}=0$ to node $i$.  Thus, $p_{j\rightarrow i}^{(1, 2)}+p_{j\rightarrow i}^{(2, 2)}=p_{j\rightarrow i}^{(1, 2)}$, so the constraints of node $i$ with respect to all other neighbor nodes are automatically removed.  The marginal probability is calculated by 

\begin{equation}
p_{i}^{c}=\frac{e^{-\beta \delta_{c}^{0}}\prod\limits_{j\in\partial i}\sum\limits_{c_{j}\in A}p_{j\rightarrow i}^{( c_{j}, c)}-(1-\delta_{ c}^{0}-\delta_{ c}^{2})\prod\limits_{j\in\partial i}\sum\limits_{c_{j}\geq c}p_{j\rightarrow i}^{(c_{j},c)}}{\sum\limits_{c_{i}}e^{-\beta \delta_{c_{i}}^{0}}\prod\limits_{j\in\partial i}\sum\limits_{c_{j}\in A}p_{j\rightarrow i}^{(c_{j}, c_{i})}-(1-\delta_{ c_{i}}^{0}-\delta_{ c}^{2})\prod\limits_{j\in\partial i}\sum\limits_{c_{j}\geq c_{i}}p_{j\rightarrow i}^{(c_{j},c_{i})}}.
\end{equation}

We implement the BPD algorithm as follows:\\
(1) Read in the given network $W$, initialize all nodes to be unobserved, and set all cavity messages $p_{i\rightarrow j}^{(c_{i},c_{j})}$ to be uniform messages. Set the inverse temperature $\beta$ to be sufficiently large (depending on the convergence of the inverse temperature). Then, iterate the BP equation using Eq. (2) until it converges to one stable point. Finally, compute the occupation probability of each node $i$ using Eq. (3).  \\
(2) Cover the small fraction $\gamma$ (e.g., $\gamma=0.01$) of unfixed nodes that have the highest covering probabilities.\\
(3) Update the state of all uncovered nodes as follows: if node $i$ is uncovered and has at least one neighbor that takes state $c_{k}=0$, then node $i$ takes state $c_{i}=1$; if node $i$ is uncovered and has at least one neighbor that takes state $c_{k}=1$, and there is no neighbor that takes state $c_{k}=0$, then node $i$ takes state $c_{i}=2$.\\
(4) Fix the state of the observed node, i.e., if the observed node $c_{i}=1$ has at most one neighbor taking the state $c_{k}=2$, then fix the state of node $i$ to $c_{i}=1$. \\
(5) If network $W$ still contains unobserved nodes, then apply the BP equation using Eqs. (2), (8), or (10). Calculate the marginal probability using Eqs. (3), (9), or (11), depending on the state of node $i$. Repeat steps (2)--(4) until all nodes are observed. 

\subsection{Greedy algorithm}
The greedy heuristic is a very simple and fast algorithm that is based on the concept of the node's impact. The impact of an uncovered node $i$ is equal to the number of nodes that will be observed by covering node $i$. For the MDS problem, we set three states for each node: $c_{i}=1$ if node $i$ is covered; $c_{i}=2$ if node $i$ is observed; $c_{i}=3$ if node $i$ is unobserved. We calculate the impact of a given uncovered node $i$ by counting the unobserved neighbor nodes in the MDS problem,  because a covered node only observes itself and its neighbor nodes in this scenario. We use $n_{i}$ to denote the number of unobserved neighbor nodes of node $i$. If node $i$ is observed, then the impact of node $i$ is equal to $n_{i}$. Otherwise, node $i$ is unobserved, and its impact is equal to $n_{i}+1$.\\
 We develop a greedy heuristic algorithm for the 2-distance MDS problem based on the concept of the node's general impact. The general impact of an uncovered node $i$ is equal to the sum of the unobserved nodes located at 2-distance from node $i$. Each node has four states in the 2-distance MDS problem: $c_{i}=0$ if node $i$ is covered; $c_{i}=1$ if node $i$ is observed and at least one neighbor node is covered; $c_{i}=2$ if node $i$ is observed and at least one 2-distance neighbor node is covered and there is no covered (1-distance) neighbor; $c_{i}=3$ if node $i$ is unobserved. In a graph without loops, the general impact of node $i$ is equal to $\sum\limits_{j\in\partial i} I_{j}-C_{i}+1$ if node $i$ is unobserved and equal to $\sum\limits_{j\in\partial i} I_{j}$ if node $i$ is observed, where $I_{j}$ denotes the impact of node $j$ and $C_{i}$ denotes the degree of node $i$. In the general case, the impact of node $i$ is equal to $n_{i1}+n_{i2}+1$ if node $i$ is unobserved and equal to $n_{i1}+n_{i2}$ if node $i$ is observed. Here, $n_{i1}$ denotes the unobserved 1-distance neighbor nodes of node $i$ and $n_{i2}$ denotes the unobserved 2-distance neighbor nodes of node $i$. \\
Starting from an input network $W$ with all the nodes unobserved, the greedy algorithm uniformly selects node $i$ at random from the subset of nodes with the highest general impact and fixes its occupation state to $c_{i}$ = 0. Then, all neighbor nodes and 2-distance neighbor nodes of node $i$ are observed. If any unobserved nodes remain in the network, then the general impact value for each of the uncovered nodes is updated and the greedy covering process is repeated until all nodes are observed. This pure greedy algorithm is very easy to implement and very fast. We found that it typically reaches a true 2-distance MDS when the input network contains a large number of edges.

\begin{figure}[htb]
  \centering
  \includegraphics[width=12cm,height=7cm]{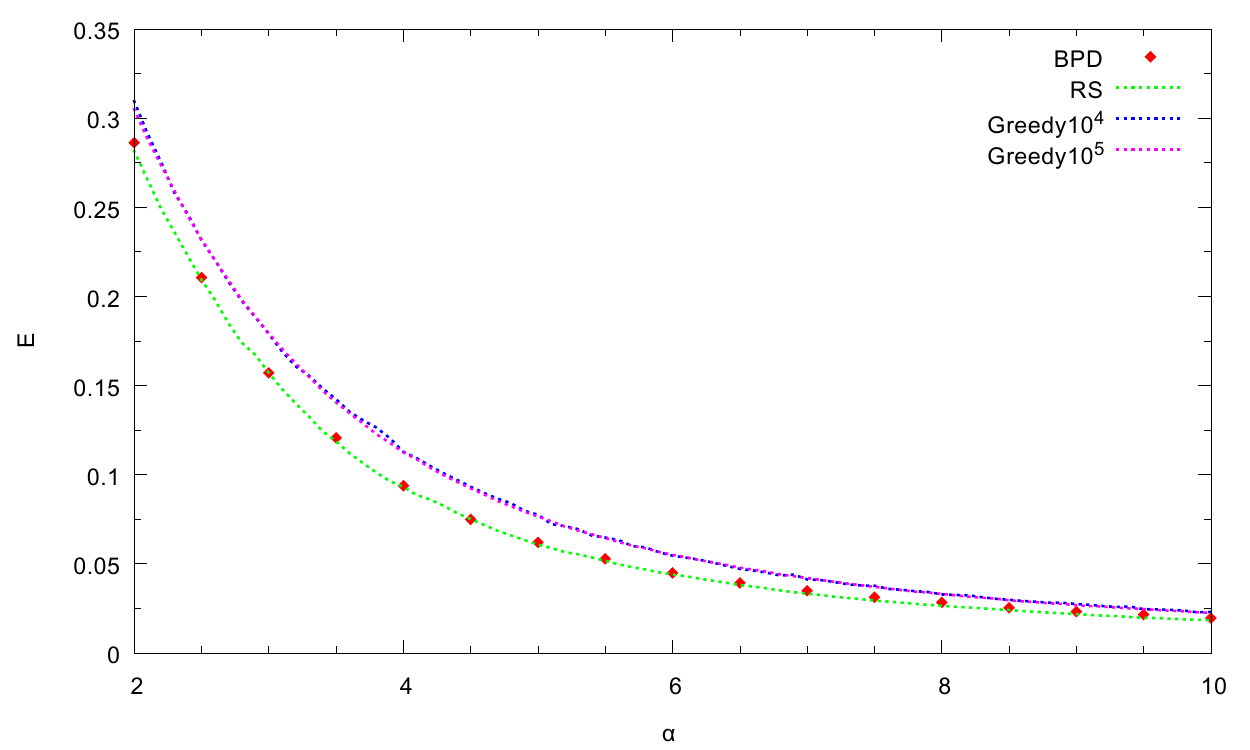}
  \caption{BPD and RS results for the 2-distance MDS problem on a single ER random graph with $N=10^{4}$ nodes, and the greedy algorithm results on single ER random graphs with $N=10^{4}$ nodes and $N=10^{5}$ nodes. The $x$-axis denotes the node degree and the $y$-axis denotes the energy density. The inverse temperature $\beta = 7.0$.}
\end{figure}
\begin{figure}[htb]
  \centering
  \includegraphics[width=12cm,height=7cm]{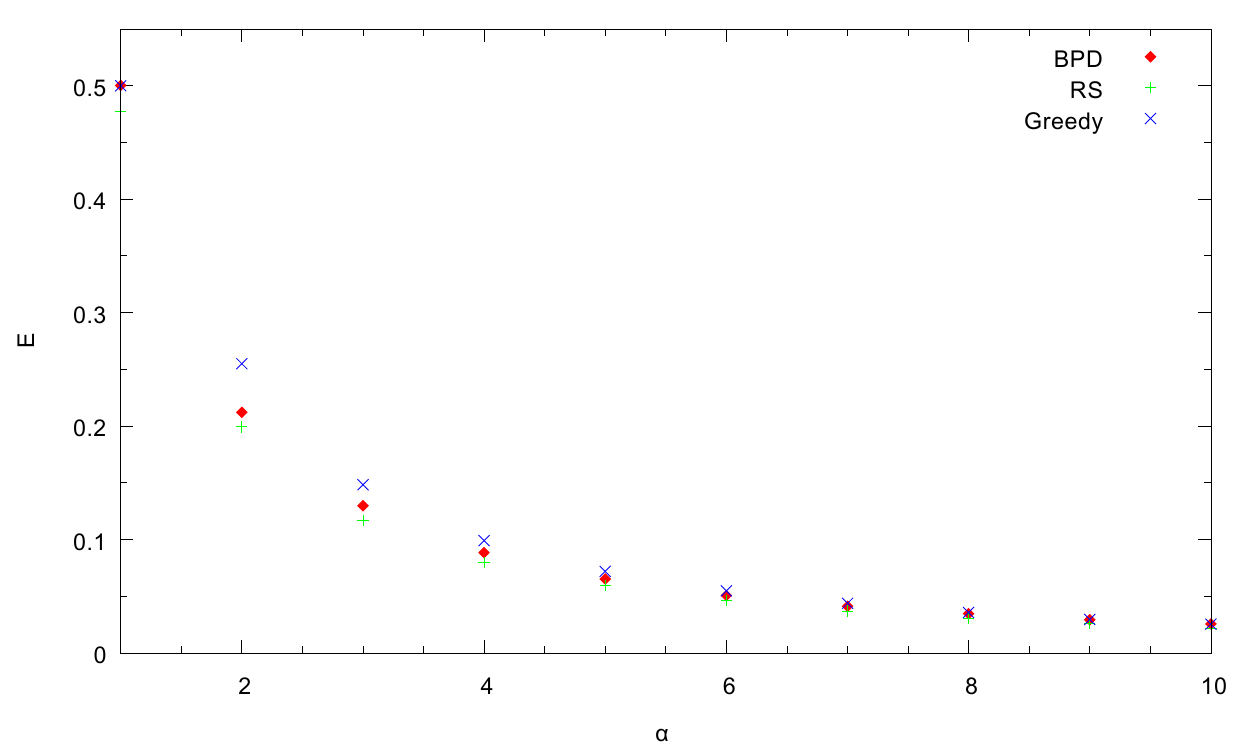}
  \caption{ BPD, greedy, and RS results for the 2-distance MDS problem on an RR random graph with $N=10^{4}$ nodes. The $x$-axis denotes the node degree and the $y$-axis denotes the energy density. The inverse temperature $\beta = 9.0$.}
\end{figure}
The results of the greedy algorithm for an ER random network and an RR network are compared with the results of the BPD algorithm in Figs. 4 and 5. The BPD algorithm outperforms the greedy algorithm, providing results that are very close to those of the RS theory on both the ER and RR random graphs. The RS results for the RR random network show that RS theory only considers the optimal graph for any node degree. Because the RR single graph contains diverse cycles,  the energy density of the BP algorithm is  greater than that given by RS theory on the single graph. However, it is almost equal to the results from RS theory when we average over many single graphs. 

\section{Rejection-free simulated annealing}
This section describes the rejection-free simulated annealing (SA) process. The SA process starts from an initial inverse temperature $\beta=\beta_{init}$, which is quite low (e.g., $\beta_{init} = 5.0$). The initial temperature is determined by the initial evolution time $\omega_{0}$, which ensures that the SA process reaches a stable energy density state at this inverse temperature. The occupation configuration $c = (c_{1} , c_{2} , . . . , c_{N})$ is initialized to be fully covered, namely $c_{i}= 0$ for all vertices $i\in G$. Each covered node contributes a unit energy, so the total energy of the initial configuration is $E(c) = N$. At each value of the inverse temperature $\beta$, configuration $c$ is allowed to evolve for a time $\omega_{0}$ through a sequence of single node flips, and the mean value of the configuration energies is recorded during this time window. The inverse temperature is then increased to $\beta\rightarrow\beta+\varepsilon$, where $\varepsilon$ is a small value, e.g., $\varepsilon = 0.001$ or $\varepsilon = 0.01$. The SA process continues to run at this and later elevated $\beta$ values until the final value $\beta_{final}$ is reached (e.g., $\beta_{final}=12.5$). The latest configuration $\textbf{c}$ is then returned as the output of the SA evolution process, and we average over 196 independent paths to obtain the average energy density. In this work, we studied both ER and RR random networks.\\
We adopt the Metropolis importance-sampling method to update the occupation configurations $\textbf{c}$. The following briefly introduces the main idea of the Metropolis importance-sampling method. We can write the energy function of the 2-distance MDS problem as
 \begin{equation}E_{i}=
\begin{cases}
1.0&\qquad c_{i}=0\qquad \sum\limits_{k\in\partial i}\delta_{c_{k}}^{2}=0\\
0.0&\qquad c_{i}=1\qquad and \qquad\sum\limits_{k\in\partial i}\delta_{c_{k}}^{0}\geq1\\
0.0&\qquad c_{i}=2\qquad and \qquad\sum\limits_{k\in\partial i}\delta_{c_{k}}^{1}\geq1,\qquad \sum\limits_{k\in\partial i}\delta_{c_{k}}^{0}=0\\
\infty&\qquad otherwise
\end{cases}
\end{equation}
The usual Monte Carlo Markov chain scheme starts from a configuration $\sigma$, and proposes a candidate move $\sigma\rightarrow \acute{\sigma}$. According to the Metropolis criterion, the acceptance probability satisfies 
\begin{equation}
P(\sigma)P(\sigma\rightarrow\acute{\sigma})=P(\acute{\sigma})P(\acute{\sigma}\rightarrow\sigma),
\end{equation}
 where $P(\sigma\rightarrow\acute{\sigma})$ represents the probability of transitioning from configuration $\sigma$ to a different configuration $\acute{\sigma}$ with some prior probability distribution $C(\sigma\rightarrow\acute{\sigma})$, and accepts the move with some acceptance rate $A(\sigma\rightarrow\acute{\sigma})$. Thus, we can write the transition probability as
 \begin{equation}
 P(\sigma\rightarrow\acute{\sigma})=C(\sigma\rightarrow\acute{\sigma})A(\sigma\rightarrow\acute{\sigma}),
 \end{equation}
  where the equilibrium probability $P(\sigma)$ is determined by the Boltzmann distribution as
 \begin{equation}
 P(\sigma)=\frac{e^{(-\beta E(\sigma))}}{z},
 \end{equation}
 which can be computed by the usual formula:
 \begin{equation}
 A(\sigma\rightarrow\acute{\sigma})=min(1,e^{-\beta(E(\acute{\sigma})-E(\sigma))}\frac{C(\acute{\sigma}\rightarrow\sigma)}{C(\sigma\rightarrow\acute{\sigma})}).
 \end{equation} 
 In a straightforward implementation, the proposed moves consist of choosing one spin uniformly at random and flipping it. The $C$ terms in the above equation are then simplified and we are left with the following simple rule:
\begin{equation}
 P(\sigma\rightarrow\acute{\sigma})=min(1,e^{-\beta(E(\acute{\sigma})-E(\sigma))}).
 \end{equation} 
 For the initial configuration $\textbf{c}$, let us denote the set of all flippable vertices from $c_{i}\in\{1, 2\}$ to $c_{i}=0$ as $V_{1\rightarrow 0}$ and the set of all flippable vertices from $c_{j}=0$ to $c_{j}\in\{1, 2\}$ as $V_{0\rightarrow 1}$. Thus, 
  \begin{equation}
 z_{\sigma}=e^{-\beta}|V_{1\rightarrow 0}| + |V_{0\rightarrow 1}|.
  \end{equation} 
 The probability of rejecting a move in a standard Metropolis scheme (Eq. (16)) is $1-z(\sigma)/N$. The rejection-free SA  procedure at each step is then as follows:\\
(1) Extract a number of iterations to skip as  $\frac{log(1-r)}{log(1-z(\sigma)/N)}$, where $r$ is a random number extracted uniformly in
$[0,1)$, and ``advance the clock'' accordingly.\\
(2) Extract a class $V_{1\rightarrow 0}$ with probability $\frac{V_{1\rightarrow 0}}{z(\sigma)}$.\\
(3) Extract a spin uniformly at random from the chosen class, and flip it.\\
(4) Update the classes.\\
\begin{figure}[!htb]
  \centering
  \includegraphics[width=12cm,height=7cm]{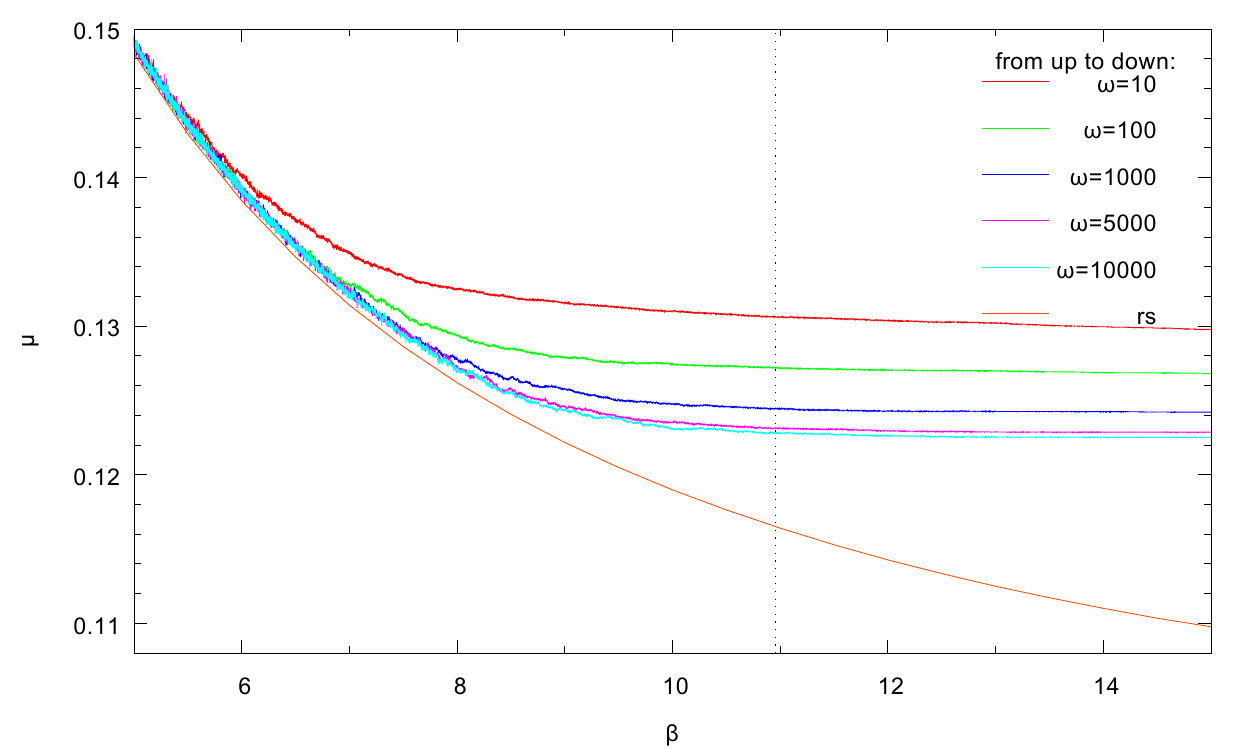}
  \caption{SA and RS results for the 2-distance MDS problem on an RR random graph with $c=3$ and $N=1000$ using different evolution times. The $x$-axis denotes the inverse temperature $\beta $ and the $y$-axis denotes the energy density.}
\end{figure}
Figure 6 shows that, if the rate of increase of $\beta$ is very fast (or if the waiting time is very small), then the energy density deviates from the value predicted by mean field theory for small $\beta$. If we slow down the rate of increase of $\beta$, then the energy density deviates from the predicted value when $\beta$ is large. The reason for this difference between SA and mean field theory is that the characteristic relaxation time lengthens as $\beta$ increases. If the characteristic relaxation time exceeds the average SA waiting time, then the configuration obtained by the SA process is not an equilibrium configuration, and the energy is higher than the average equilibrium energy.\\

\section{Discussion}
In this paper, we have proposed a greedy-impact local algorithm and a BPD message-passing algorithm and have developed the RS mean field theory for solving the network 2-distance dominating set problem algorithmically and theoretically. We found that the BP equation always converges to a stable point in RR networks when the mean degree is from 3--9 (see Fig. 3), and converges in ER random networks when the mean degree is from 4.2--10.4; the equation cannot converge when the inverse temperature exceeds a certain threshold in other circumstances (see Fig. 2).  The existence of the minimum energy limit has been proved using simulated annealing on RR networks. The solution space of the 2-distance MDS has a different structure on ER and  RR networks according to the mean degree. The one-step RS breaking theory could be used to study the solution space of the 2-distance MDS problem. Our numerical results (Figs. 4 and 5) suggest that the mean-field BPD algorithm constructs a near-optimal 2-distance MDS for random networks, and that the mean-field BPD algorithm is better than the greedy algorithm.\\
A direct extension of our work is to consider the 2-distance MDS problem for directed networks. We have been work on the directed 2-distance MDS problem in this year\cite{44}. A great deal of theoretical work remains to be studied. A more challenging and common problem in dominating sets is the connected dominating set problem. We will use spin glass theory [25] to study both the minimal connected dominating set problem  and the 2-distance minimal connected dominating set problem.  We will also study various types of dominating set problem (e.g., double dominating set [1,5], liar's dominating set [2], and extended dominating set [4]) using spin glass theory.  
\section*{$\hspace{2mm}$ Acknowledgements}
Yusupjan Habibulla thanks Prof. Haijun Zhou for helpful discussions and guidance. This research was supported by the doctoral startup fund of Xinjiang University of China (grant number 208-61357) and partially supported by the National Natural Science Foundation of China (grant numbers 11765021, 11705279, and 61662078). We thank Maxine Garcia, PhD, from Liwen Bianji, Edanz Group China (www.liwenbianji.cn/ac) for editing the English text of a draft of this manuscript.

\end{document}